# Round-Robin Test of a Light-Emitting Electrochemical Cell: Establishing a Reference Protocol for Quality Research

Anton Kirch[1], Kumar Saumya[1], Joan Ràfols-Ribé[1], Shi Tang[1], Christian Larsen[1], Ajay Kumar Poonia[1,2], Nicolò Maccaferri[1,2,12], Chang-Ki Moon[3], João Pedro Ferreira Assunção[4], Frank Nüesch[4], Sandra Gellner[5], Rubing Bai[6], Weiao Yang[7], Zuowei Liu[7], Daniel Tordera[8], Sergio Martínez-Saiz[8], Shun-ichiro Ito[9], Koshi Oi[9], Felix Hergenhan[10], Karl Sebastian Schellhammer[10], Sebastian Reineke[10], Taishi Takenobu[9], Henk J. Bolink[8], Yufeng Hu[7], Zhiwei Liu[6], Ekaterina Nannen[5], Roland Hany[4], Malte C. Gather[3,11], and Ludvig Edman[1,12]*

* Correspondence: ludvig.edman@umu.se

[1] The Organic Photonics and Electronics Group, Department of Physics, Umeå University, SE-90187 Umeå, Sweden

[2] Ultrafast Nanoscience Group, Department of Physics, Umeå University, SE-90187 Umeå, Sweden

[3] Humboldt Centre for Nano- and Biophotonics, Department of Chemistry and Biochemistry, University of Cologne, Greinstr. 4-6, 50939 Cologne, Germany

[4] Empa, Swiss Federal Laboratories for Materials Science and Technology, Laboratory for Functional Polymers, Überlandstrasse 129, CH-8600 Duebendorf, Switzerland

[5] Competence Centre for Novel Electronics and Advanced Materials, Faculty of Engineering and Computer Science, University of Applied Sciences Niederrhein Krefeld, Germany

[6] Beijing National Laboratory for Molecular Sciences, State Key Laboratory of Rare Earth Materials Chemistry and Applications, College of Chemistry and Molecular Engineering, Peking University, Beijing, 100871, P. R. China

[7] Key Laboratory of Luminescence and Optical Information, Ministry of Education, Institute of Optoelectronic Technology, Beijing Jiaotong University, Beijing, 100044, P.R. China

[8] Instituto de Ciencia Molecular (ICMol), University of Valencia, C/ Catedrático José Beltrán 2, 46980 Paterna (Valencia), Spain

[9] Department of Applied Physics, Nagoya University, Furo-cho, Chikusa, Nagoya 464-8603, Japan

[10] Dresden Integrated Center for Applied Physics and Photonic Materials (IAPP) and Institute of Applied Physics (IAP), Technische Universität Dresden, Dresden, Germany

[11] Organic Semiconductor Centre, SUPA School of Physics and Astronomy, University of St Andrews, St Andrews KY16 9SS, United Kingdom

[12] Wallenberg Initiative Materials Science for Sustainability, Department of Physics, Umeå University, SE-90187 Umeå, Sweden

**Abstract**

Emerging technologies benefit from a jointly established “reference protocol”, which can lower the bar of entry for new researchers while serving as a calibration standard for established actors. The light-emitting electrochemical cell (LEC) combines electrochemistry and optoelectronics in an intricate manner, and it can by that enable sustainable and commercially relevant printing fabrication of emissive thin-film devices. However, LEC performance is sensitive to a range of material and processing parameters, which frequently results in inadequate, or even erroneous, device evaluation. With this in mind, we present herein a LEC reference protocol, which details the sourcing of materials and the procedures and parameters for robust device fabrication and operation. The protocol has been tested across nine international research groups, and the collected results from this interlaboratory round-robin test confirm that good LEC performance can be reproducibly obtained following our protocol. We also identify common pitfalls that can arise during LEC development, and present practical steps for attaining optimum LEC performance. We hope this reference protocol will improve the quality of future LEC research and serve as a guide for future researchers entering this vibrant field.

## Main

The light-emitting electrochemical cell (LEC) was invented in the mid-1990s by Qibing Pei, Alan J. Heeger, and co-workers at the University of California at Santa Barbara.[1] It essentially comprises a blend of an electroluminescent organic semiconductor (OSC) and mobile ions in a single-layer active material (AM) sandwiched between two electrodes. The first LEC used a fluorescent conjugated polymer as the OSC,[1] but subsequent studies have demonstrated that it is possible to employ phosphorescent ionic[2] and neutral[3, 4] transition metal complexes, fluorescent small molecules,[5] quantum dots,[6] and organic compounds that emit by thermally activated delayed fluorescence (TADF)[7, 8] and multi-resonance TADF[9] as the emitter. The mobile ions in the AM commonly originate from either an alkaline salt dissolved in a polar solid-state solvent,[1, 3] an ionic liquid,[10] or the "free" ions that compensate ionic groups chemically grafted onto an ionic OSC.[2]

When a voltage is applied between the two electrodes, the mobile ions drift towards the electrode interfaces, where they form thin electric double layers (EDLs). This EDL-formation continues until facile and balanced electron and hole injection is accomplished. The first injected electrons and holes cause additional redistribution of the mobile ions by virtue of their space charge, with cations migrating to compensate the injected electrons for the formation of n-type doping, while the anions compensate the injected holes for p-type doping. These growing p- and n-type regions eventually make contact to form a p-n (or p-i-n) junction in the AM. At this junction, subsequently injected electrons and holes efficiently recombine into excitons, which can radiatively decay under the emission of photons. This LEC "turn-on" process continues until all mobile ions are either part of the EDLs or contribute to the electrochemical doping[11, 12] and it is manifested in a lowering of the drive voltage (if the LEC is driven in constant-current mode) and an increase in the luminance with time.

This unique operational mode renders the LEC technology practically relevant, since it is possible to utilize solely air-stable and soluble device materials and a robust device architecture for the accomplishment of efficient electron and hole injection, transport, and recombination, and thereby efficient exciton generation at low voltage. More specifically, it is possible to fabricate *complete* functional LECs by sustainable and scalable printing and coating processes, with demonstrated methods including slot-die[13] and spray[14] coating. This versatile fabrication opportunity also enables the direct conformable coating of LEC devices onto complex surfaces and substrates, such as fibers,[15] textiles[16] and even forks[14].

However, it is important to also acknowledge that the LEC performance is highly sensitive to a wide range of material, processing, and operational parameters. For instance, the time to reach significant luminance for an LEC depends on the ionic conductivity of the AM, which in turn is determined by its mobile ion concentration and nano-morphology,[17, 18] and on the biasing protocol and history[19]. The emission efficiency of LECs is strongly influenced by both the position[20] and the width[21] of the p-n junction, which are intricately dependent on the AM thickness,[20] the electronic structure and the electron and hole mobility of the OSC(s),[22] the mobile ion concentration,[23, 24] and the biasing protocol[19]. Finally, the stability and efficiency of LEC devices can be severely influenced by the selection of electrodes[25] or AM constituents[26], or the accidental inclusion of contaminants[27] that cause electrochemical or chemical side reactions.

This complexity means that it is easy to make performance-degrading mistakes during the design, fabrication, and operation of LECs, with unfortunate consequences being that promising materials and device designs are discarded prematurely and that new researchers are scared away from developing an exciting emerging technology. The ambitions of this round-robin test are thus to effectively address these issues by introducing a straightforward and robust LEC "reference protocol", by verifying its validity through its independent execution at nine international research laboratories, and by pinpointing common mistakes that should be avoided during LEC development to enable a fair material and device analysis and an optimum device performance.

**Results**

The latest version of the reference protocol is presented in Note S1 in the Supporting Information. It describes in detail the sourcing and pretreatment of the constituent materials, substrates, and solvents, the preparation of ink solutions, the device fabrication procedure, and the execution of the measurements. This protocol was originally developed at the Umeå University node during more than one decade of LEC research, and its first version (see Note S2) was distributed to the nine participating research groups in the spring of 2025.

The research groups were instructed to not only follow the protocol and report on their collected measurement data, but also to communicate on any forced or accidental deviation from the protocol and their specific observations and recommendations that could improve the protocol for future users. Note S1 presents the updated version of the reference protocol that has considered this input. The specific instrumentation and procedures employed by the participating research groups are presented in Note S3. The identity of each research group has herein been anonymized, and the data are presented as being contributed by "Lab A", "Lab B", etc.  A summary and discussion of the key steps in the reference protocol, including key insights and suggestions from the different research groups, follows below.

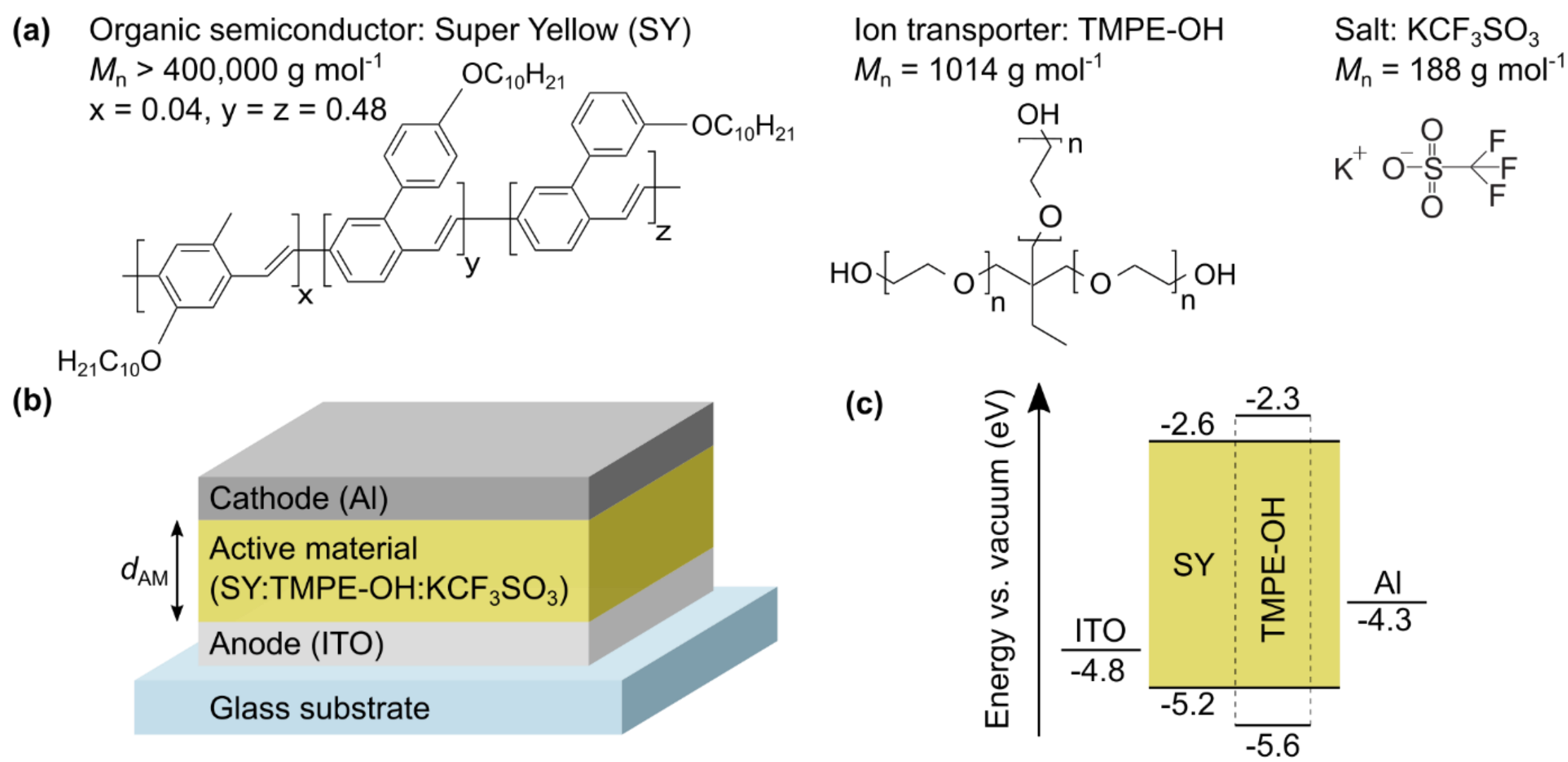

***Figure 1.*** *(a) The active-material constituents, (b) the LEC device architecture, and (c) the frontier energy levels of key device constituents.*[25, 28]

Figure 1(a) displays the chemical structure of the three constituents in the AM: the electroluminescent OSC in the form of the conjugated co-polymer "Super Yellow" (SY, $M_n$ > 400.000 g mol$^{-1}$), the ion-dissolving and ion-transporting compound hydroxyl-capped trimethylolpropane ethoxylate (TMPE–OH, $M_n$ = 1014 g mol$^{-1}$), and the salt $KCF_3SO_3$ (purity ≥ 99.5%). More details regarding vendor information, CAS number, etc. can be found in Note S1.

We emphasize that the materials in this study were selected for the straightforward attainment of a repeatable and robust LEC performance and commercial availability, and not for the achievement of records. In this context, we do however mention that more efficient LEC emission obviously can be obtained with an electroluminescent OSC that can convert triplet excitons to photons (e.g. a triplet or TADF emitter), that a slightly better LEC performance has been attained with a lower molecular weight of,[29] or a different end-capping unit on,[30] the TMPE-based ion transporter, and that Lab G reports on that the employment of $LiCF_3SO_3$ instead of $KCF_3SO_3$ for the salt results in a slightly higher emission efficiency but at the cost of a slower turn-on time and poorer device repeatability.

TMPE-OH and $KCF_3SO_3$ are dried in a vacuum oven at $T$ = 50 °C for ≥ 12 h and $T$ = 190 °C for ≥ 16 h, respectively, before being transferred, together with the as-received SY, into an inert-atmosphere glovebox for ink formulation and device fabrication. The three AM constituents are separately dissolved in cyclohexanone (purity ≥ 99.5%) in a solute concentration of 8 g l$^{-1}$ for SY and 10 g l$^{-1}$ for TMPE-OH and $KCF_3SO_3$. The three master solutions are stirred on a hot plate at $T$ = 70 °C for ≥ 24 h. It is recommended that at least 5 ml, but preferably 10 ml, of each master solution is prepared to minimize the weighing error. Moreover, it is critical that the time of stirring of the SY solution is extended until SY is observed to be fully dissolved. Important, we have identified that a source of contamination and black-spot formation during LEC operation and storage are impurities in the $KCF_3SO_3$ salt, but that this issue can be rectified by filtering the $KCF_3SO_3$ solution through a standard poly(tetrafluoroethylene) (PTFE) filter, with a pore size of 0.1 µm.

The three master solutions are mixed to form the "AM-ink" with a SY:TMPE-OH:$KCF_3SO_3$ mass ratio of 1:0.1:0.03. Since the SY solution is highly viscous – and therefore difficult to accurately pipette – it is recommended that the mixing of the master solutions is executed by adding the required volumes of the TMPE-OH and $KCF_3SO_3$ solutions to a pre-weighed SY solution. The AM-ink is finally stirred on a hot plate at $T$ = 70 °C for ≥ 6 h. Note that the prepared AM-ink should not be filtered, since this might result in an alteration of the AM mass ratio (which can have a strong influence on the LEC performance).[24, 31]

***Table 1.*** *A summary of key device and performance parameters, as reported by the nine different Labs. The LEC devices were driven by a constant current density of 10 mA cm$^{-2}$.*

| *Lab* | *$d_{AM}$ (nm)* | *Fabrication Yield* | *Lifetime Yield* | *$V_{min}$ ± σ (V)* | *$L_{peak}$ ± σ (cd m$^{-2}$)* | *$CE_{peak}$ ± σ (cd A$^{-1}$)* | *$LE_{peak}$ ± σ (lm W$^{-1}$)* |
|---|---|---|---|---|---|---|---|
| *A* | 160 ± 5 | 8/8 | 8/8 | 3.1 ± 0.1 | 350 ± 30 | 3.5 ± 0.3 | 3.2 ± 0.3 |
| *B* | 92 ± 13 | 8/8 | 8/8 | 3.8 ± 0.6 | 300 ± 10 | 3.0 ± 0.1 | 2.3 ± 0.4 |
| *C* | 104 ± 4 | 8/8 | 8/8 | 3.0 ± 0.1 | 430 ± 20 | 4.3 ± 0.2 | 4.4 ± 0.2 |
| *D* | 140 | 8/8 | 1/8 | 4.4 ± 0.5 | 990 ± 230 | 9.9 ± 2.3 | 2.9 ± 1.7 |
| *E* | 108 ± 5 | 8/8 | 8/8 | 2.8 ± 0.1 | 440 ± 40 | 4.4 ± 0.4 | 4.7 ± 0.4 |
| *F* | 92 ± 8 | 18/22 | 18/18 | 3.1 ± 0.2 | 340 ± 60 | 3.4 ± 0.6 | 3.1 ± 0.6 |
| *G* | 90 ± 10 | 8/8 | 8/8 | 3.0 ± 0.1 | 470 ± 40 | 4.7 ± 0.4 | 4.6 ± 0.2 |
| *H* | 96 ± 12 | 6/8 | 3/6 | 5.3 ± 0.7 | 400 ± 190 | 4.0 ± 1.9 | 2.5 ± 1.3 |
| *I* | 74 ± 11 | 8/8 | 6/8 | 4.8 ± 0.9 | 270 ± 70 | 2.7 ± 0.7 | 1.8 ± 0.8 |

The AM-ink is thermally equilibrated to room temperature, and thereafter spin-coated (acceleration = 2000 rpm s$^{-1}$, speed = 2000 rpm, time = 60 s) onto a cleaned ITO-coated glass substrate, with the following specifications: ITO sheet resistance ≤ 20 Ω sq.$^{-1}$, optical transparency (at $\lambda$ = 555 nm) > 85 % and glass area = 3-10 cm$^2$. After spin-coating, the sample is immediately transferred to a hot plate where the AM film is dried at $T$ = 70 °C for 1 h. We emphasize that, since the important thickness and morphology of the dry AM film are dependent on the spin-coating and drying procedure, the above instructions should be followed exactly to attain optimum device performance. In this context, we advise verifying that the hot-plate temperature is displayed correctly (see Figure S1).

A set of reflective Al top electrodes is deposited on top of the dry AM film through a shadow mask by thermal evaporation under vacuum. It is strongly recommended to protect the AM film with a shutter during the first few seconds of evaporation until a steady Al deposition rate is reached. Moreover, it is advised to use a low deposition rate of 1-3 Å s$^{-1}$, until at least a continuous (~20 nm thick) Al film has formed, to avoid damage to the soft AM film below. The spatial overlap of the bottom ITO and the top Al electrodes defines the area and the number of the emissive LEC pixels on each glass substrate, which in the round-robin test varied between 3 and 6 $mm^2$ and 4 and 16, respectively.

Figures 1(b,c) are a schematic of the essential LEC device structure and the energy levels of key device constituents, respectively, with $d_{AM}$ (in Fig. 1b) defining the dry thickness of the AM film. Table 1, first column, presents the reported average value (and the standard deviation if several values were provided) for $d_{AM}$ by each group. This metric varies quite significantly from 74 nm for Lab I, over a typical value of ~90-110 nm for a majority of the Labs, to reach 140 and 160 nm for Labs D and A, respectively.

A deviation from the typical $d_{AM}$ value can originate from one or several of the following issues: (i) the formulation of the AM-ink, with, e.g., a higher SY concentration than the nominal resulting in a larger $d_{AM}$; (ii) the spin-coating execution, notably the employment of incorrect spin-coating velocity or acceleration, a hot AM-ink (with lowered viscosity but faster solvent evaporation), or a time delay between ink deposition and spinning onset; (iii) the measurement of $d_{AM,}$ by probing a non-representative region, e.g., the material overshoot next to a scratch or the edge-bead region that is characteristic of spin-coated films.

The participating research groups were instructed to fabricate at least eight nominally identical LECs and to characterize them either inside a high-quality inert-atmosphere glove box or to encapsulate them with a glass slide for evaluation under ambient conditions. We further emphasize that the research groups were instructed to only measure on pristine devices, i.e. LECs that had not been biased previously. The implementation of this “pristine LEC” requirement is strongly recommended as the basis for any reporting of LEC data, since the “slow” relaxation of bias-induced mobile-ion concentration gradients in the LEC active material otherwise can result in very-difficult-to-analyze, or even misleading, device metrics.

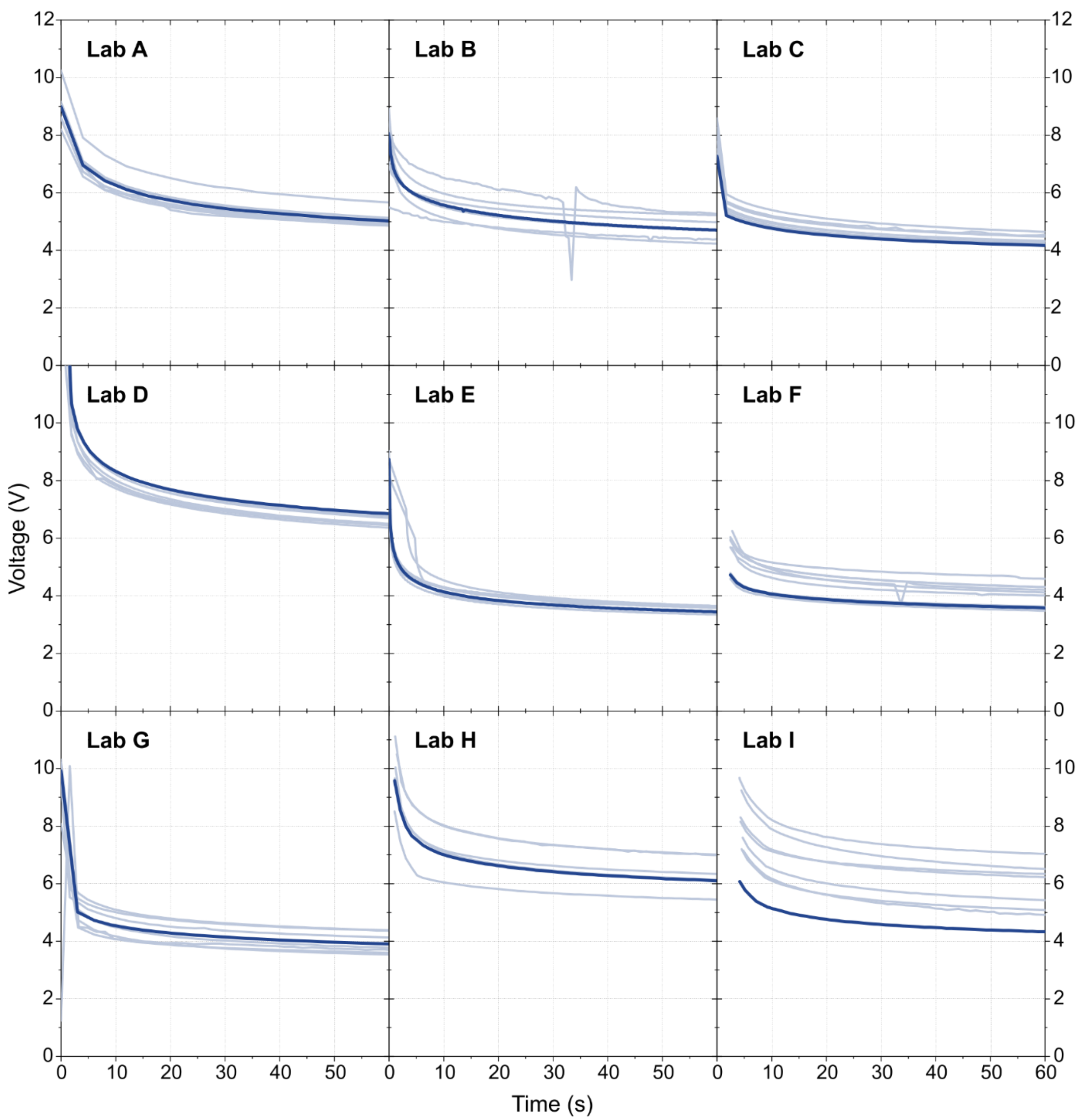

***Figure 2.*** *The initial temporal evolution of the drive voltage of the functional pristine devices from the nine research groups ("Labs") during driving by a constant current density of 10 mA cm$^{-2}$. The median-performing devices (in the long-term luminance measurement) are highlighted with a darker color.*

The pristine LEC devices are driven by a constant current density of $j$ = 10 mA cm$^{-2}$ and with the voltage compliance set to 21 V. The temporal evolution of the voltage and the luminance is measured for at least 20 h, with the sampling time interval preferably being less than 2 s during the first 60 s of measurement ("the turn-on measurement") and ≤ 1 min for the remainder of the measurement ("the long-term measurement"). Six of the research groups performed the turn-on and the long-term measurements on the same set of pristine devices, whereas the other three groups measured on two different sets of devices.

Table 1, second column, presents the reported "fabrication yield", which is defined as the ratio between the number of "functional devices" and the number of fabricated devices. A device is defined to be "functional" if, during the long-term measurement, it: (i) delivers a peak luminance above 100 cd $m^{-2}$, and (ii) exhibits a drive voltage of ≥2.5 V for at least 1 h of continuous operation. The robustness of the reference-LEC architecture is confirmed by that seven of the nine research groups reported a fabrication yield of 100 %, while the other two (Labs F and H) still achieved a reasonable fabrication yield of ~80 %.

Figure 2 presents the time evolution of the drive voltage for all functional LEC devices (separated into different graphs for each research group or Lab) during the turn-on measurement. The measurement curves highlighted in darker color in Figures 2-5 represent the device that delivered the median value for the peak luminance during the long-term measurement. The consistent observation for all 80 devices is that the voltage decreases during the initial operation. This initial voltage drop is characteristic of functional LEC devices, since the initial redistribution of the mobile ions in the AM of a functional LEC improves both the injection and the bulk transport of the electronic charge carriers.

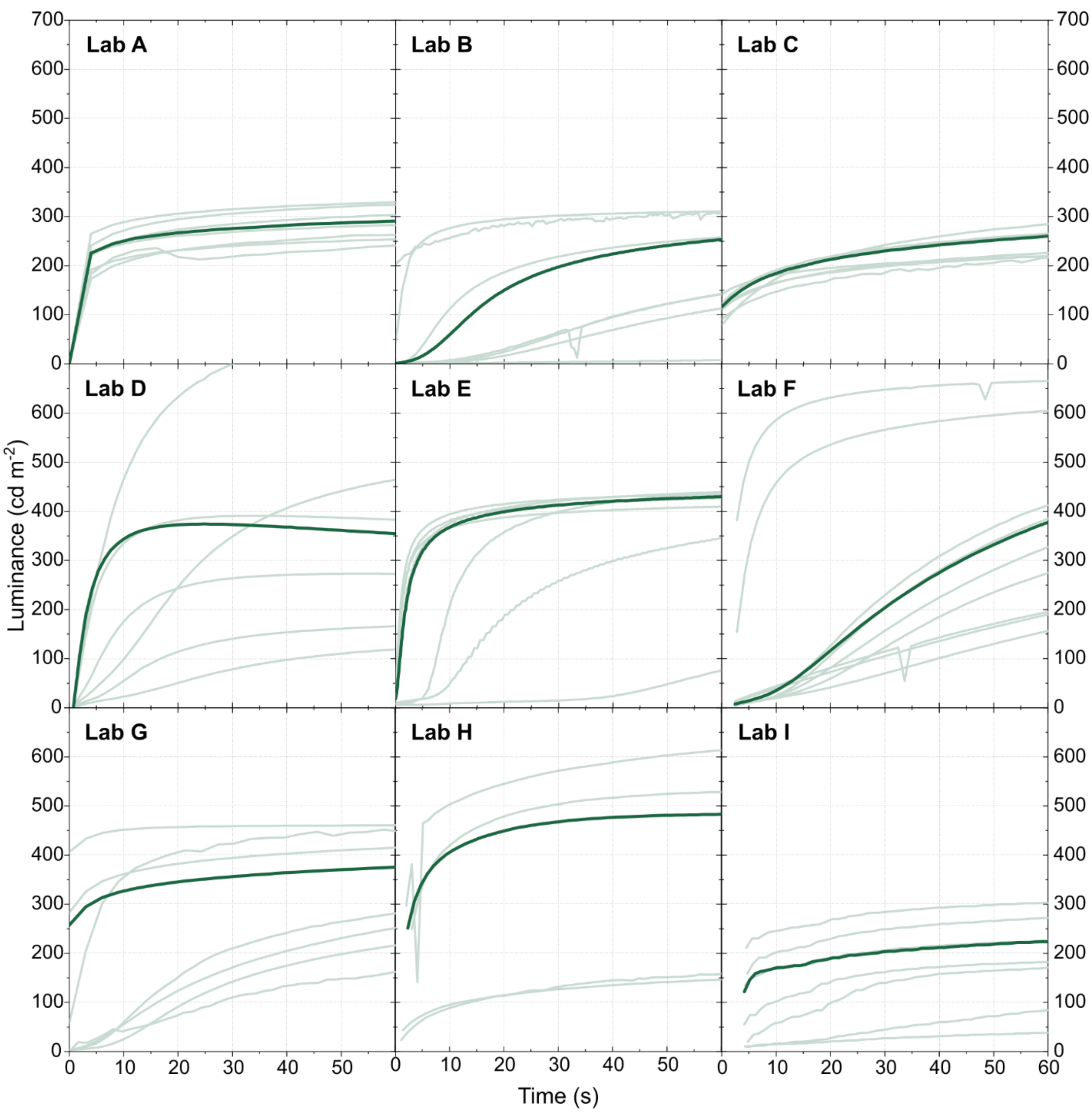

***Figure 3.*** *The initial temporal evolution of the luminance of the functional pristine LEC devices from the nine laboratories during driving by a constant current density of 10 mA cm$^{-2}$. The darker line indicates the median device in the long-term luminance measurement.*

Figure 3 displays the corresponding luminance turn-on transients for the functional devices of each lab, with the consistent observation being that the luminance increases significantly during the first 1 min of constant-current operation. This finding is important since it shows that the initial ion migration in the AM results in the formation of EDLs at the electrode interfaces, which enable *efficient and balanced* hole and electron injection into SY. Thereby, the exciton formation rate and luminance are increased, despite the existence of high electron and hole injection barriers at the two electrode interfaces (see Fig. 1c).

An admittedly critical drawback with many LEC systems from an application viewpoint is a slow turn-on time, i.e. a problematically long time delay between the application of electric bias and the attainment of significant luminance. The LEC turn-on time is primarily determined by the ionic conductivity of the AM (although other factors, such as bias protocol[19] and doping redistribution during operation,[32] also can have an influence). The fact that all nine research groups in this study reported at least two pristine LEC devices that reach a luminance of 100 cd $m^{-2}$ within only a few seconds of operation shows that the appropriately fabricated AM film exhibits a sufficiently high ionic conductivity to render the reference LEC practically relevant for applications.

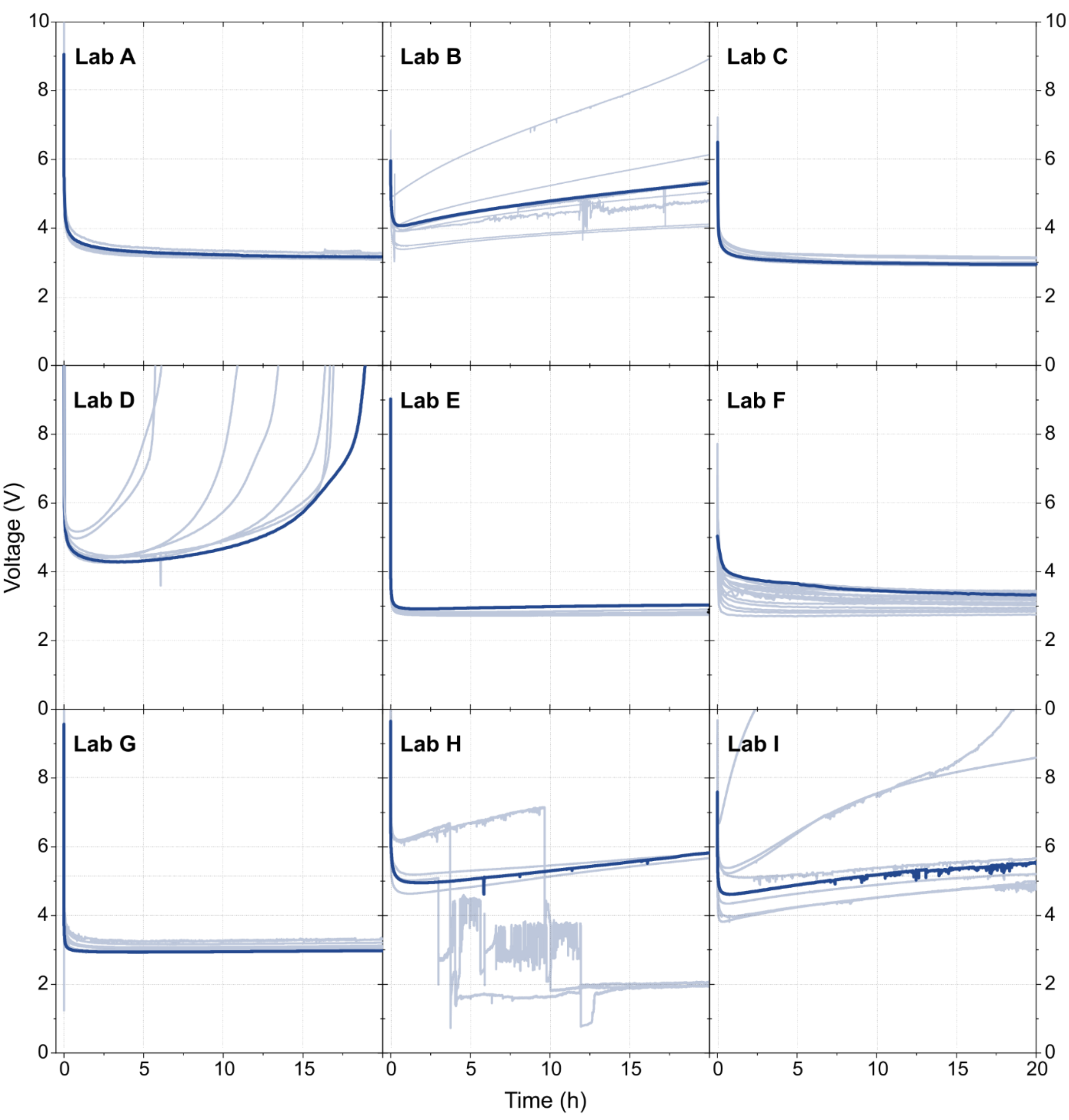

***Figure 4.*** *The long-term evolution of the drive voltage of the functional LEC devices from the different labs. The pristine devices were driven by a constant current density of 10 mA cm$^{-2}$. The darker line indicates the median device in the long-term luminance measurement.*

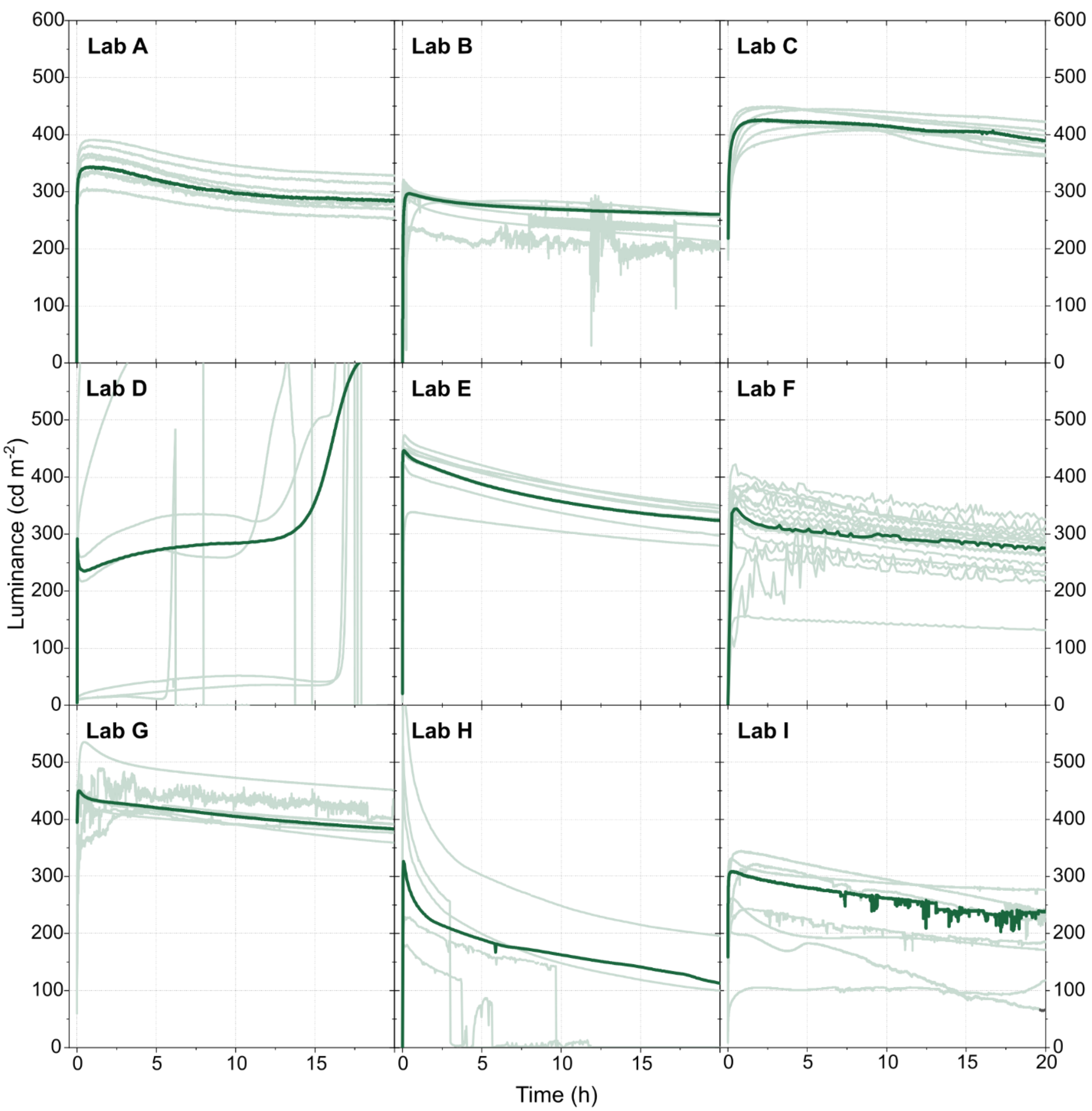

***Figure 5.*** *The long-term evolution of the luminance of the functional LEC devices from the different labs. The pristine devices were driven by a constant current density of 10 mA cm$^{-2}$. The darker line indicates the median device in the long-term luminance measurement.*

Figures 4 and 5 display the evolution of the drive voltage and the forward luminance, respectively, of the pristine LEC devices during the long-term measurement, as reported by the nine research groups. Figure S2 presents the electroluminescence (EL) spectrum recorded after 1 min, 1 h, and 20 h during this long-term measurement. The observation that all nine research groups report on an EL spectrum that is rather time invariant suggests that cavity effects induced by a moving emission zone (a phenomenon that can be common in LEC devices)[32] are negligible and can be excluded from the analysis below.

A qualitative inspection of the device data in Figures 4 and 5 yields that five research groups (Labs A, C, E, F, and G) report comparable voltage and luminance performance for all their devices during the entire 20 h period of measurement, whereas the other four labs report on devices that suffer from stability and repeatability issues. We mention that "well-behaved" LEC operation was reported for both non-encapsulated devices operated in a glovebox (Labs C, E, and G) and for encapsulated devices operating under ambient air (Labs A and F).

We begin our quantitative stability analysis by defining the "lifetime yield" as the quotient between the number of stable devices and the number of functional devices, with a "stable" device exhibiting a luminance exceeding 100 cd $m^{-2}$ and a drive voltage above 2.5 V for at least 20 h of continuous operation. Table 1 shows that the lifetime yield for the well-behaved devices from Labs A, C, E, F, G (but also from Lab B) is 100 %, whereas it is much lower for Labs D, H, and I.

More specifically, the well-behaved LECs (from Labs A, C, E, F and G) feature a low minimum voltage ($V_{min}$) of ~3.0 V, with a very minor overall device-to-device variation of 0.1-0.2 V (see Fig. 4 and Table 1). The peak luminance ($L_{peak}$) for these well-behaved LECs also exhibits a small device-to-device variation within the same research group, while the interlaboratory variation for the average value of $L_{peak}$ is larger, as it ranges from 340 cd $m^{-2}$ (Lab F) to 470 cd $m^{-2}$ (Lab G).

In this context, we note that the research groups employed different procedures for the measurement of the forward luminance, with the solid collection angle spanning from a low value of <0.01 sr to full forward-hemisphere collection at $2\pi$sr. Although the reference-LEC exhibits close-to-Lambertian emission (see Figure S3), this variation in collection angle will affect the recorded luminance. It is further possible that small variations in the AM fabrication can contribute, since it has been demonstrated that the emission performance of similar LEC devices is highly sensitive to both the concentration of mobile ions in their AM[12] and to $d_{AM}$[33, 34] (see Figure S4).

The peak current efficacy ($CE_{peak}$) is derived by dividing $L_{peak}$ with the current density, while the peak luminous efficacy ($LE_{peak}$) is the maximum for the quotient between the luminance and the product of the current density, the drive voltage, and $\pi$ (i.e. under the assumption of Lambertian emission). Table 1 shows that $CE_{peak}$ for the well-behaved LEC devices varies between 3.4 and 4.7 cd $A^{-1}$, while $LE_{peak}$ ranges from 3.1 to 4.6 lm $W^{-1}$, which are good values, albeit not records, for fluorescent LECs.[24, 30]

We now shift our attention to the analysis of the lower-performing LECs in Figures 4 and 5. Lab D reports LEC devices with low voltage and luminance stability, as quantified by the small lifetime yield of 1/8. These devices end their operational lifetime by the drive voltage reaching compliance (Fig. 4) and the luminance exhibiting an anomalously high value for a very short time before quickly dropping to zero (Fig. 5). In consideration of that Lab D reported that they "suspect that air may have leaked into the test fixture [during the long-term measurement]", it is reasonable to assign this poor performance to the existence of water and/or oxygen in the AM, which causes detrimental electrochemical side reactions. It has, for instance, been established that the existence of water in the AM of an LEC can result in an electrochemical water-splitting reaction at the negative Al cathode, with the generated $H_2$ gas causing delamination of Al from the AM.[27] This gradual loss of functional device area will result in a concomitantly increased electric power dissipation in the remaining functional device regions, when both the voltage and the effective current density increase. This increased dissipation of electric power in a decreasing functional device volume can result in the formation of candescent hot spots that would then be the origin of the measured transient "luminance" peaks.

Labs B and I report qualitatively similar problematic behavior in the form of a voltage increase and a luminance decrease, albeit at a lower rate than Lab D. The fact that the rate of the voltage increase is more rapid than the luminance drop suggests that the degradation is primarily taking place at the electrode interfaces (or in the doped regions) and not in the emissive p-n junction. It thus seems as though also this performance drop is due to a contaminant-induced electrochemical side reaction, although the exact origin of the contaminant is unclear.

Finally, half of the functional devices reported by Lab H exhibit a correlated, distinct drop in voltage and luminance, while the remaining functional devices suffer from a marked luminance reduction during the long-term measurement. Since the same Lab reported the lowest fabrication yield (see Table 1), it seems plausible that the culprit could be an accidental inclusion of contaminant particles into the AM film, which renders it sensitive to the formation of short circuits during (and before) LEC operation.

**Discussion**

The collected results of this round-robin test confirm that it is indeed possible to realize LECs with reproducible performance by following the reference protocol (for the updated version, see Note S1). All nine participating research groups reported on a high fabrication yield of 75-100% and a short turn-on time of a few seconds to a brightness exceeding 100 cd $m^{-2}$. Five of the research groups further reported repeatable emission performance over the assigned measurement period of 20 h for their entire set of 8 devices. This promising performance can be quantified by the following typical metrics: a minimum voltage of 3.0 V, a peak luminance of 400 cd $m^{-2}$ at a current efficacy of 4.2 cd $A^{-1}$, and a peak luminance efficacy of 4.0 lm $W^{-1}$.

However, it should also be acknowledged that four of the participating research groups experienced different setbacks during the longer-term electric driving of their devices, which resulted in a markedly lower device performance. We therefore consider it appropriate to present a summary of our general view on culprits that can cause subpar LEC performance and how these issues can be avoided or rectified.

First, it is critical to select the "non-active" LEC materials – for the reference LEC: the two electrodes, the ion transporter, and the salt – so that they remain electrochemically stable or inert during LEC operation. This implies that the negative (positive) electrode is inert at the potential at which the OSC(s) is n-type (p-type) doped, whereas the salt and the ion transporter must remain inert within the entire potential range that is spanned by the electrochemical p- and n-type doping potentials of the OSC(s). The reference LEC is designed with this in mind, since its Al and ITO electrodes have been demonstrated[25] to be electrochemically stable at the n-type and p-type doping potentials of SY, respectively, whereas the $KCF_3SO_3$ salt and the TMPE-OH ion transporter are electrochemically stable within the potential window spanned by the n- and p-type doping potentials of SY.[29] A facile electrochemical method for evaluating whether this electrochemical stability criterion is fulfilled can be found in References [[25, 29]].

Second, the AM must be free from "reactive impurities" when the LEC is driven to light emission. The existence of such impurities can otherwise cause electrochemical side reactions at the electrode interfaces because of the high electric fields in the EDLs and chemical side reactions in the bulk of the AM through interactions with energetic polarons and excitons. Impurities that must be removed from the AM of an LEC before it is driven to light emission include water, oxygen, and solvent residues, as well as impurities originating from low-purity compounds (e.g. the $KCF_3SO_3$ salt). In the reference protocol, the first three impurities are removed by drying the AM film in a vacuum oven, whereas the removal of a known non-volatile impurity is done by filtration. A direct consequence of this requirement is that the AM must be protected from ambient air by an appropriate barrier during LEC operation. It should also be noted that larger dust particles should not be apparent during the device fabrication, since they can damage the thin AM film and cause short circuits.

Third, the composition and thickness of the AM need to be carefully tuned for the realization of high-performance LEC operation. The turn-on time to light emission is dependent on the ionic conductivity, which implies that faster turn-on kinetics can be attained by increasing the mobile-ion concentration or by improving the ion mobility by adding more ion transporter. However, it has also been shown that the mobile-ion concentration determines the (average) doping concentration at steady state, and that an increase in the doping concentration in an LEC is concomitant with an effective thinning of the emission zone (EZ) and increased exciton quenching.[35, 36] Thus, the take-home message is that the AM composition should be optimized to simultaneously realize fast turn-on and high emission efficiency. Furthermore, the light emission in LECs originates from a typically thin emission zone (essentially defined by the size of the p-n junction), with the consequence being that emission-altering cavity effects can be prominent in LEC devices. The reference LEC is designed with an AM composition and $d_{AM}$ for the reduction of detrimental exciton-quenching and cavity effects and for the attainment of a good balance between fast turn-on kinetics and efficient and stable emission.

To conclude, the LEC is an emerging technology that offers both exciting (and complex) science through the in-situ formation of a p-n junction doping structure and potential for practical applications since it can be cost-efficiently and sustainably fabricated by printing. We thus consider it timely to now jointly establish an LEC reference protocol, which can serve as a field standard, while also pinpointing common mistakes that should be avoided to fairly and accurately evaluate new materials and device concepts. We herein investigate the merit of our proposed reference protocol in an interlaboratory round-robin test that comprises nine different research groups, and verify that it is indeed possible to realize a robust and repeatable LEC performance by appropriately following the protocol. We hope that the presentation and verification of the reference protocol will catalyze and improve the quality of future LEC research, as well as serving as a guideline and inspiration for future researchers that consider entering this vibrant field.

## Methods

Details about material selection and processing as well as device fabrication and characterization are presented in Note 1 in the Supplementary Information.

## Data availability

All experimental data are available for download at: post link after acceptance

## Acknowledgements

L.E. acknowledges financial support from the European Union through an ERC Advanced Grant (ERC, InnovaLEC, 101096650). A.K. acknowledges funding from the European Union (HORIZON MSCA 2023 PF, acronym UNID, grant number 101150699). A.K.P. and N.M. acknowledge financial support from the Swedish Research Council (Grants No. 2021-05784 and No. 2025-04734), the Knut and Alice Wallenberg Foundation (Grant No. 2023.0089), and Kempestiftelserna (WISE-UmU/Kempe-program, Grant No. JCSMK 23-198). C.M. and M.C.G acknowledge financial support by the Alexander von Humboldt Foundation (Humboldt-Professorship to M.C.G.) and the European Union (ERC Advanced Grant, acronym HyAngle, grant number 101097878). J.P.F.A, F.N. and R.H. acknowledge financial support from the Swiss National Science Foundation (Grant No. CRSII5_216629/1). S.G. and E.N. acknowledge financial support from the German Research Foundation (Deutsche Forschungsgemeinschaft (DFG), project ID 498131727) and thank Julia Demmer for her valuable support during the setup of the laboratory equipment and for helpful discussions throughout the project. R.B. and Z.L. acknowledge financial support from the National Natural Science Foundation of China (22525101). D.T., S.M.S. and H.B. acknowledge support from the Ministry of Science and Innovation (MCIN) and the Spanish State Research Agency (AEI): project PID2024-159087OB-I00 funded by MICIU/AEI/10.13039/501100011033" and "ERDF A way of making Europe". F.H. thanks Dr. Karolis Leitonas and Thomas Tarnow for fruitful discussions and assistance during the fabrication of the LECs. S.R. acknowledges funding by the Federal Ministry of Research, Technology and Space (BMFTR) within the project "Organische IR-Emitter mit zyklischen Liganden - beeIR" (13N17176) and by the European Union (ERC, SLOWTONICS, 101089234). Y.H., W.Y. and Z.L acknowledge financial support from the National Natural Science Foundation of China (No.62275013). T.T. acknowledges financial support from JST CREST (Grant Number JPMJCR23A4).

# Supplementary Information:

## Round-Robin Test of a Light-Emitting Electrochemical Cell: Establishing a Reference Protocol for Quality Research

Anton Kirch[1], Kumar Saumya[1], Joan Ràfols-Ribé[1], Shi Tang[1], Christian Larsen[1], Ajay Kumar Poonia[1,2], Nicolò Maccaferri[1,2,12], Chang-Ki Moon[3], João Pedro Ferreira Assunção[4], Frank Nüesch[4], Sandra Gellner[5], Rubing Bai[6], Weiao Yang[7], Zuowei Liu[7], Daniel Tordera[8], Sergio Martínez-Saiz[8], Shun-ichiro Ito[9], Koshi Oi[9], Felix Hergenhan[10], Karl Sebastian Schellhammer[10], Sebastian Reineke[10], Taishi Takenobu[9], Henk J. Bolink[8], Yufeng Hu[7], Zhiwei Liu[6], Ekaterina Nannen[5], Roland Hany[4], Malte C. Gather[3,11], and Ludvig Edman[1,12]*

* Correspondence: ludvig.edman@umu.se

[1] The Organic Photonics and Electronics Group, Department of Physics, Umeå University, SE-90187 Umeå, Sweden

[2] Ultrafast Nanoscience Group, Department of Physics, Umeå University, SE-90187 Umeå, Sweden

[3] Humboldt Centre for Nano- and Biophotonics, Department of Chemistry and Biochemistry, University of Cologne, Greinstr. 4-6, 50939 Cologne, Germany

[4] Empa, Swiss Federal Laboratories for Materials Science and Technology, Laboratory for Functional Polymers, Überlandstrasse 129, CH-8600 Duebendorf, Switzerland

[5] Competence Centre for Novel Electronics and Advanced Materials, Faculty of Engineering and Computer Science, University of Applied Sciences Niederrhein Krefeld, Germany

[6] Beijing National Laboratory for Molecular Sciences, State Key Laboratory of Rare Earth Materials Chemistry and Applications, College of Chemistry and Molecular Engineering, Peking University, Beijing, 100871, P. R. China

[7] Key Laboratory of Luminescence and Optical Information, Ministry of Education, Institute of Optoelectronic Technology, Beijing Jiaotong University, Beijing, 100044, P.R. China

[8] Instituto de Ciencia Molecular (ICMol), University of Valencia, C/ Catedrático José Beltrán 2, 46980 Paterna (Valencia), Spain

[9] Department of Applied Physics, Nagoya University, Furo-cho, Chikusa, Nagoya 464-8603, Japan

[10] Dresden Integrated Center for Applied Physics and Photonic Materials (IAPP) and Institute of Applied Physics (IAP), Technische Universität Dresden, Dresden, Germany

[11] Organic Semiconductor Centre, SUPA School of Physics and Astronomy, University of St Andrews, St Andrews KY16 9SS, United Kingdom

[12] Wallenberg Initiative Materials Science for Sustainability, Department of Physics, Umeå University, SE-90187 Umeå, Sweden

**Note S1: The Updated Reference-LEC Protocol**

*1. Active-material constituents*

The active material (AM) comprises three constituents:

1. The fluorescent conjugated polymer Super Yellow (SY, Merck, $M_n$ >400.000 g mol$^{-1}$, CAS: 26009-24-5)
2. The ion-transporter hydroxyl-capped trimethylolpropane ethoxylate (TMPE-OH, Merck, $M_n$ = 1014 g mol$^{-1}$, CAS: 50586-59-9)
3. The salt potassium trifluoromethanesulfonate ($KCF_3SO_3$, Solvionic, purity ≥99.5%, CAS: 2926-27-4)

SY is used as received from the vendor. TMPE-OH is dried in a vacuum oven at $T$ = 50 °C for ≥ 12 h, and $KCF_3SO_3$ is dried in a vacuum oven at $T$ = 190 °C for ≥ 16 h at pre-vacuum conditions (< $10^{-3}$ bar). All three AM constituents (SY after delivery, and TMPE-OH and $KCF_3SO_3$ after drying) are thereafter stored in an inert-atmosphere glove box ([$O_2$], [$H_2O$] < 1 ppm) before further processing to prevent contamination.

*2. Substrate*

The indium-tin-oxide (ITO) coated glass substrate features the following metrics: ITO sheet resistance ≤ 20 Ohm/square, optical transparency ≥ 85 % at $\lambda$ = 555 nm, glass area = 3-10 cm$^2$, glass thickness ~1 mm. The ITO-coated glass substrate is cleaned by sequential 10-min sonication in: (i) detergent in deionized (DI) water, (ii) DI water, (iii) acetone, and (iv) isopropanol. The cleaned substrate is dried at $T$ > 100 °C for ≥ 1 h in an oven, and thereafter protected from dust, by e.g. coverage with a glass-cover lid, until the spin-coating of the AM-ink.

*3. Master solutions*

Three vials are cleaned by sonication in acetone, followed by drying at $T$ > 100 °C for ≥ 1 h in an oven. A clean magnetic stirring rod of appropriate size (i.e. with its long axis slightly smaller than the inner diameter of the vial) is inserted into each vial. All subsequent preparation steps are preferably performed in an inert-atmosphere glove box ([$O_2$], [$H_2O$] < 1 ppm).

Three master solutions with cyclohexanone (purity ≥ 99.5%, Merck, CAS: 108-94-1) as the solvent are prepared. The amount of cyclohexanone added to each vial is at least 5 ml, but preferably 10 ml, to decrease the weighing uncertainty. The solution volume should not exceed 3/4 of the vial volume to ensure proper stirring. The solute concentrations are:

1. SY: 8 g l$^{-1}$
2. TMPE-OH: 10 g l$^{-1}$
3. $KCF_3SO_3$: 10 g l$^{-1}$

The three master solutions are stirred on a hot plate at $T$ = 70 °C for ≥ 24 h.

Note 1: The time for dissolution of SY depends on the SY grain density and may take significantly longer than 24 h. Thus, make sure that SY is completely dissolved before moving to the next step. This can be probed by shaking the vial lightly and observing for optical irregularities in the solution, which then indicates non-dissolved SY. The dissolution of SY can also be probed by immersing a fresh needle into the solution, and to check for the existence of a floating blob of undissolved SY polymer on the solution.

Note 2: The $KCF_3SO_3$ master solution shall, after the stirring but before the formulation of the AM ink, be filtered with a 0.1 µm PTFE filter (Cytiva, Puradisc 25, Whatman) to remove impurities.

*4. AM-ink*

The three master solutions are mixed to form the AM-ink with a SY:TMPE-OH:$KCF_3SO_3$ mass ratio of 1:0.1:0.03.

Note: This is the solute (and not the solution) mass ratio.

Since the SY solution is very viscous (and therefore essentially impossible to accurately pipette), the procedure is that the appropriate volumes of the TMPE-OH and $KCF_3SO_3$ master solutions are added to the SY master solution. The AM-ink is stirred on a hot plate at $T = 70$ °C for ≥ 6 h, and thereafter stored in the inert-atmosphere glove box before further processing.

Note 1: The AM-ink must not be filtered since this may change the mass ratio and thereby the LEC performance.

Note 2: The stirring of the AM-ink should not be too vigorous, since this will result in the formation of gas cavities in the AM-ink.

Note 3: The AM-ink can be stored for an extended amount of time, provided that it is stored in an inert atmosphere (e.g. a glovebox) and in darkness.

*5. AM film fabrication*

The AM-ink (at $T \approx 20$ °C) is deposited on top of the ITO-coated substrate so that it covers the entire substrate area, and thereafter spin-coated at 2000 rpm (acceleration = 2000 rpm $s^{-1}$) for 60 s. The spin-coated AM film is protected from dust by a cover lid and dried on a hot plate at $T = 70$ °C for 1 h.

Note 1: Do not dry the AM film for neither a longer nor a shorter time, since this can have a significant influence on the LEC performance.

Note 2: It is recommended to confirm that the surface temperature of the hot plate is at 70 °C with a thermometer.

The thickness of the dry AM film can be determined by e.g. a stylus profilometer, an atomic force microscope, or an ellipsometer. The first two methods require scratching the dried AM film with some sort of sharp stylus.

Note 1: It is recommended to measure the AM film thickness at several spots close to the area where the LEC is to be fabricated to obtain an accurate estimate of the thickness and uniformity of the AM film.

*6. Top electrode fabrication*

The Al top electrode should preferably be deposited immediately after the drying of the AM film, but at the latest within 12 h. The Al electrode is deposited on top of the AM film to a thickness of 100 nm by thermal evaporation at a base pressure of $< 5\times10^{-6}$ mbar. An evaporation shutter shall protect the AM film during the initial Al evaporation until a steady evaporation rate is reached. It is advised to use a low deposition rate of 1-3 Å $s^{-1}$, until at least a continuous (~20 nm thick) Al film has formed, to avoid damage of the soft AM film below.

*7. Device characterization*

The characterization must be performed on pristine (i.e. non-biased) LEC devices, and it should be executed within 24 h of their fabrication. This measurement can either be executed on non-encapsulated LEC devices inside an inert-atmosphere glove box ([$O_2$], [$H_2O$] < 1 ppm) or on appropriately glass-encapsulated devices under ambient air. We advise to measure a larger set (> 8) of LEC devices to establish the device-to-device variation. We also recommend to report on the emissive LEC pixel area (as defined by the spatial overlap of the Al and ITO electrodes), the substrate area, and the number of emissive LEC pixels on each substrate.

The pristine LEC devices are driven by a constant drive current density of $j$ = 10 mA $cm^{-2}$, with ITO biased as the positive anode and with the voltage compliance set to 21 V. It is required to report on the collection angle in the luminance measurement. The temporal evolution of the voltage and the luminance is measured during > 20 h, with the sampling time interval preferably being ≤ 2 s during the first 60 s of measurement, and ≤ 1 min for the remainder of the measurement. Alternatively, one set of pristine LEC devices is measured with the higher sampling rate of ≤ 2 s for 60 ("the turn-on measurement"), while a second set of pristine LEC devices is measured at the lower sampling rate of ≤ 1 min for > 20 h ("the lifetime measurement").

The EL spectrum is measured on (at least) one pristine LEC device after ~1 min, ~1 h, and ~20 h of operation at $j$ = 10 mA $cm^{-2}$. The detection angle for the EL spectrum must be reported. This spectral measurement can either be performed in parallel with an already ongoing V-L-t measurement or on a separate pristine LEC device.

**Note S2: The First Version of the Reference-LEC Protocol that was Communicated to the Participating Research Groups**

*1. Material purchase and preparation*

The active material comprises three constituents, purchased from Sigma-Aldrich and Solvionic:

1. The emissive polymer Super Yellow (SY)
   (Sigma-Aldrich, CAS: 26009-24-5)
2. The ion transporter hydroxyl-capped trimethylolpropane ethoxylate (TMPE-OH)
   (Sigma-Aldrich, CAS: 50586-59-9, $Mn$ = 1014 g/mol)
   - If the expected delivery time is too long, we can provide this material
3. The salt potassium trifluoromethanesulfonate ($KCF_3SO_3$)
   (Solvionic, purity ≥ 99.5%, CAS: 2926-27-4)

Use SY as received. Dry TMPE-OH in a vacuum oven at 50 °C for ≥ 12 h. Dry $KCF_3SO_3$ in a vacuum oven at 190 °C for ≥ 16 h. Store the dried TMPE-OH, the dried $KCF_3SO_3$, and SY in a glove box to prevent water contamination.

*2. Substrate cleaning*

Use glass substrates with an indium tin oxide (ITO) coating (sheet resistance ≈ 20 Ohm/square; transparency at 555 nm ≥ 85 %). Clean the substrates by subsequent sonication in (i) detergent in DI water, (ii) DI water, (iii) acetone, and (iv) isopropanol for ≥ 10 min each. Dry them above 100 °C for ≥ 1 h in an oven. Make sure that the substrates are covered and protected from dust after cleaning.

*3. Master ink preparation*

Clean 3 vials by sonication in acetone followed by drying in an oven. Insert one clean stirring rod into each vial. All the following ink preparation and fabrication steps should take place in an inert-atmosphere glove box with oxygen and water levels preferably below 1 ppm (if not, please specify the levels).

Prepare three master inks with cyclohexanone (≥ 99.5% purity, Sigma Aldrich, CAS: 108-94-1) as the solvent. Each ink quantity is > 5 ml to decrease weighing uncertainties. The solute concentrations are:

1. SY: 8 g/l
2. TMPE-OH: 10 g/l
3. $KCF_3SO_3$: 10 g/l

Stir the three master solutions on a hot plate at 70 °C for ≥ 24 h. Make sure that the SY is completely dissolved. This may take sometimes significantly longer than 1 day depending on the SY grain densities.

*4. Active-material ink preparation*

The active-material ink has a mass ratio (SY:TMPE-OH:$KCF_3SO_3$) of 1:0.1:0.03. As the SY solution is very viscous and does not allow for accurate pipetting, the active-material ink is prepared by adding the respective amounts of the TMPE-OH and $KCF_3SO_3$ solutions to the SY solution. Stir the active-material ink on a hot plate at 70 °C for ≥ 6 h. Store the active-material ink in a glove box and use it preferably within two weeks after preparation.

*5. Spin-coating active material*

Let the active-material ink cool down to room temperature before spin coating. Deposit a sufficient amount of ink onto the ITO-coated substrate to cover its entire area, and spin coat at 2000 rpm (acceleration = 2000 rpm/s; ramp time = 1 s) for 60 s. Dry the active-material films on a hot plate at 70 °C for 1 h, using a cover lid to protect the films from dust.

NOTE 1: Do not dry the film for a longer or shorter time, since this may affect the device performance.

NOTE 2: Ensure that the hot plate surface temperature is at 70 °C using a thermometer.

*6. Top electrode evaporation*

The top electrodes should be fabricated preferably immediately after the drying of the active-material film, at the latest after 12 hours. Deposit a 100 nm thick layer of aluminum by thermal evaporation at a base pressure $<5\times10^{-6}$ mbar. Make sure to use an evaporation shutter to not deposit aluminum oxide.

*7. Device characterization*

IMPORTANT: Only use pristine devices!

The LECs must be characterized within 24 h after fabrication either unencapsulated inside a glove box or encapsulated using a cover glass outside the box (please report back).

Measure the temporal evolution of the forward luminance and the voltage over > 20 hours using a constant drive current density of $j$ = 100 A/m$^2$ (= 10 mA/cm$^2$) and a voltage compliance of 21 V. The ITO is biased as the positive anode. Specify your pixel area.

The sampling interval should be ≤ 2 s during the first 60 s of the measurement. For the remainder of the measurement, it should be ≤ 1 min. If these specifics cannot be met, a set of samples can be measured with a low sampling rate for > 20 h ("lifetime measurement") and another set of nominally identical samples can be measured with a high sampling rate for a short period ("turn-on measurement").

Please specify the geometry that is used for the forward luminance measurement, i.e. detection angle, distance LEC-detector, and detection area.

For at least one sample, three forward EL spectra are to be recorded at about 1 min, 1 h, and 20 h after turning on the device, using the same driving condition ($j$ = 10 mA/cm$^2$ ) as above. Specify the detection angle for the EL spectra measurement. Depending on the experimental setup, this measurement can be performed on a separate, nominally identical sample or while measuring the *j*-*V*-*L* transients.

*9. Active-material thickness*

The thickness of the active material is measured on one of the dried films without a top electrode at three different spots on a pixel area, using, e.g., a profilometer like Bruker Dektak.

*10. Number of samples*

Every group prepares and measures at least 8 nominally identical (same person(s) fabricating, all made from the same ink) devices and records at least 8 *j*-*V*-*L* transients as specified above. If the "lifetime" and "turn-on" characteristics cannot be recorded in the same measurement, then at least 8 samples should be measured for each characteristic.

IMPORTANT: All device data must be reported. For instance, if one or more devices are non-functional (e.g. shorted or shifty), they must still be part of the report.

The EL spectra can be recorded on an extra sample if the parallel recording of the EL spectra during the *j*-*V*-*L* measurement is not possible.

The sample that is used for the film thickness measurement is not to be used for device fabrication, because it has been handled outside the glove box.

NOTE: We recommend that test batches are fabricated and evaluated prior to the batch of samples that is to be reported. The persons fabricating and evaluating the samples should be familiar with the procedure and have gained some experience to rule out common failures on the "first attempt".

*11. Data evaluation and presentation*

After receiving the raw data, we will anonymize the group names and label the data sets as "lab 1, …, lab *N".* We will present and compare the median of the *j*-*V*-*L* curves of the functional devices (with respect to the peak forward luminance) of the working samples of all labs in one plot, probably one sub-figure per lab. One figure will show the turn-on behavior (zoom into the first minute of operation) and a second figure the 20 h lifetime measurement. Further, we will calculate and present the yield (100% - x % of dysfunctional samples), the measured active-material thickness, and the EL spectra for each lab. The minimum voltage, peak forward luminance, peak PCE (assuming Lambertain emission), and their respective standard deviations for each lab will be presented in a Table. We will present all data (in one *j*-*V*-*L* graph for each group) in the SI. All data will be published alongside the manuscript.

**Note S3: Methods and Instrumentation Reported by the Participating Research Groups**

*Lab A*

Device fabrication

Indium tin oxide (ITO)-coated (thickness = 100 nm, Rs < 20 Ω $sq^{-1}$), glass substrates (substrate area = 24 x 24 $mm^2$, thickness = 0.7 mm, Xinyan Technology Ltd., HK). Eight LEC emissive pixels were fabricated on each ITO-coated glass substrate, with the emissive pixel area being 2 x 2 $mm^2$.

Device characterization

The AM thickness ($d_{AM}$) was measured with a stylus profilometer (Dektak XT, Bruker, USA) by scratching the AM layer in the sample middle and measuring it at eight different positions. A glass cover was attached to each device using a UV-curable NOA 68 resin (Norland Products) for encapsulation. The voltage-luminance transients were recorded on devices with glass-lid encapsulation by a lifetime test system (M6000, McScience Inc.). The LECs were electrically driven and measured by a computer-controlled source-measure unit (Keithley 2450, Keithley Instruments), with the voltage compliance set to 21 V. The luminance was measured with a calibrated photodiode (PDA100A2, Thorlabs) and read by a multimeter (Keithley 2100, Keithley Instruments). The PD photoactive area was 75.4 $mm^2$, and the distance between the LEC and the PD's active area is about 168 mm (solid collection angle of about 0.0027 sr). Sampling interval during measurement: 4 s during the first 60 s, and 54 s for 60 s < t < 20 h. The EL spectra were measured with a calibrated spectrometer (Ocean HDX, Ocean Insight) with the fiber head (QP600-2-VIS-BX, 600 µm core diameter) positioned 225 mm away from the LEC (solid collection angle of about 5.6×10-6 sr).

The samples were not exposed to air at any point before/during the measurement due to the hermetic encapsulation inside the glovebox (see b above).

*Lab B*

Device fabrication

Indium tin oxide (ITO)-coated (thickness = 150 nm, Rs = 10 Ω $sq^{-1}$, transparency of ≥ 85% at 555 nm) glass substrates (substrate area = 20 x 20 $mm^2$, thickness = 0.7 mm, (Geomatec Co., Ltd., Japan)). Four emissive LEC pixels were fabricated on each ITO-coated glass substrate, with the emissive pixel area being 2.0 x 2.0 $mm^2$.

Device characterization

The AM thickness ($d_{AM}$) was measured with an AFM (FlexAFM 5, Nanosurf, Switzerland) by scratching the AM layer in the sample middle and measuring it at four different positions.

The voltage-luminance transients were recorded on non-encapsulated devices inside a N2-filled glovebox ([O2], [H2O] < 1 ppm, T ≈ 25 °C).

The LECs were electrically driven and measured by a semiconductor parameter analyzer (B1505A, Agilent Technologies, Inc.), with the voltage compliance set to 21 V. The luminance was measured with a calibrated photodiode (PD; S1337-21, Hamamatsu Photonics) and read out directly by a semiconductor parameter analyzer (B1505A, Agilent Technologies, Inc.). The PD photoactive area was 2.0 x 2.0 $cm^2$, and the distance between the LEC and the PD's active area is about 0.7 mm.

Sampling interval during measurement: 0.5 s

The EL spectra were measured using a calibrated spectrometer (SA-100C-HPCB/C, Lambda Vision), where the emission was collected and focused through a microscope objective lens (LMPlan IR 5×, Olympus Corp.).

*Lab C*

Device Fabrication

Indium tin oxide (ITO)-coated (thickness = 100 nm, Rs = 20 Ω $sq^{-1}$) glass substrates (substrate area = 24 × 24 $mm^2$, thickness = 0.7 mm, South China Science & Technology Company Limited). Six LEC emissive pixels were fabricated on each ITO-coated glass substrate, with the emissive pixel area being 2 × 2 $mm^2$.

Device Characterization

The AM thickness ($d_{AM}$) was measured with an Alpha-Step IQ Surface Profiler by scratching the AM layer in the sample middle and measuring it at six different positions. The voltage-luminance transients were recorded on non-encapsulated devices inside a $N_2$-filled glovebox ([O2] < 5 ppm, [H2O] < 1 ppm, T ≈ 22 °C). The LECs were electrically driven and measured by a computer-controlled source-measure unit (Keithley 2400), with the voltage compliance set to 21 V. The EL spectra, current, voltage, luminance, and external quantum efficiency (EQE) were simultaneously measured using an absolute EQE measurement system (C9920-12, Hamamatsu Photonics) coupled with a photonic multichannel analyzer (PMA-12, Hamamatsu Photonics). The system was equipped with an integrating sphere accessory, in which the LECs under test were mounted on the inner wall of the sphere (solid collection angle of about 2π sr).

Sampling interval during measurement: 1.7 s for the first 60 s, then 1 min for t > 60 s.

All fabrication, storage, and measurement processes for the LECs were conducted in an $N_2$-filled glovebox. The only exception was the transfer (≤ 30 s) between the fabrication and measurement gloveboxes, which was carried out using a sealed, $N_2$-filled transfer vessel to avoid air exposure.

*Lab D*

Device fabrication

ITO-coated glass substrates (sheet resistance = 15 Ω/sq, dimensions = 16.35 × 15.95 $mm^2$, thickness = 1.1 mm; Shenzhen South China Xiangcheng Technology Co., Ltd.). Six LEC pixels were prepared on each substrate, with each pixel area measuring 3 × 1.5 $mm^2$. Note: The AM ink was spin-coated at room temperature.

Device characterization

The active material thickness ($d_{AM}$) was measured with a stylus profilometer (Ambios Technology XP-2, USA) by scratching the active layer in the center of the sample and taking measurements at four different positions.

Voltage–luminance (V–L) transients were measured with the test fixture. The device was encapsulated into the fixture inside a nitrogen glovebox ([$O_2$], [$H_2O$] < 1 ppm, T ≈ 27 °C). After encapsulation was completed within the glovebox, the fixture assembled with the encapsulated device was transferred out of the glovebox for subsequent V–L transient testing. During the long-term (20 h) lifetime test, we suspect that air may have leaked into the test fixture. This could explain the voltage increase observed in the later stage of the measurement.

The LECs were driven by a computer-controlled source meter (Keithley 2450) with the voltage compliance set to 21 V. The luminance was measured with a silicon photodetector (Hamamatsu S2281, active area = 100 $mm^2$) connected to a multimeter (Keithley 2010) via a current-to-voltage amplifier. The detector was placed about 5 mm from the device. The sampling interval during the lifetime measurement was 40 s.

The EL spectra were collected with a calibrated spectrometer (Ocean Optics UV-3101PC) using a fiber probe (core diameter = 1000 µm) positioned approximately 5 mm from the LEC.

*Lab E*

Device fabrication

Indium tin oxide (ITO)-coated (thickness = 145 nm, Rs = 20 Ω sq-1) glass substrates (substrate area = 30 x 30 mm2, thickness = 0.7 mm, Kintec, HK). Four LEC pixels were fabricated on each ITO-coated glass substrate, with the emissive pixel area being 2 x 2 $mm^2$.

Device characterization

The AM thickness (*d*AM) was measured with a stylus profilometer (Dektak XT, Bruker, USA) by scratching the AM layer in the sample middle and measuring it at four different positions.

The voltage-luminance transients were recorded on non-encapsulated devices inside a N2-filled glovebox ([O2], [H2O] < 1 ppm, T ≈ 22 °C).

The LECs were electrically driven and measured by a computer-controlled source-measure unit (Agilent U2722A), with the voltage compliance set to 21 V. The luminance was measured with a calibrated photodiode (PD) equipped with an eye-response filter (S9219-01, Hamamatsu Photonics), connected to a data acquisition card (National Instruments USB-6009) via a current-to-voltage amplifier, and read by the same SMU. The PD photoactive area was 3.6 x 3.6 mm2, and the distance between the LEC and the PD's active area is about 12 mm (solid collection angle of about 0.1 sr).

Sampling interval during measurement: 100 ms during the first 2 s, 200 ms for 2 s < t < 10 s, 500 ms for 10 s < t < 60 s, 1 s for 60 s < t < 120 s, 5 s for 120 s < t < 300 s, 10 s for 300 s < t < 1 h, 1 min for t > 1 h

The EL spectra were measured with a calibrated spectrometer (USB2000+, Ocean Optics) with the fiber head (QP600-2-VIS-BX, 600 µm core diameter) positioned about 5 mm away from the LEC (solid collection angle of about 0.01 sr).

*Lab F*

Device fabrication

Indium tin oxide (ITO)-coated (thickness = 90 nm, $R_s$ = 20 Ω $sq^{-1}$) glass substrates (substrate area = 25 x 25 $mm^2$, thickness = 1 mm, Suzhou ShangYang Solar TechnologyCo., Ltd). Four LEC emissive pixels were fabricated on each ITO-coated glass substrate, with the emissive pixel area being 6.35 $mm^2$.

The devices were encapsulated using an epoxy resin (XNR5516Z-L and XNR5590, Nagase Europa GmbH) inside the glovebox. The encapsulation glass (Sodalime glass, AMGTECH Korea) comprises a small cavity above the pixels that prevents direct contact with the sensitive materials.

Device characterization

The AM thickness ($d_{AM}$) was measured with a stylus profilometer (Veeco Dektak 150) by scratching the AM layer in the sample middle and measuring it at four different positions. The voltage-luminance transients were recorded on encapsulated devices outside the glovebox at room temperature. Due to limitations of available measurement setups, the turn-on and lifetime measurements were carried out in two different measurement setups.

Turn-on: The LECs were electrically driven and measured by a computer-controlled source-measure unit (SMU-2450, Keithley Instruments Inc.), with the voltage compliance set to 21 V. The luminance was measured in a calibrated integrating sphere (LMS-100, Labsphere Inc.) with a photodiode (PD). The PD was calibrated by determining the luminance with a spectrometer (CDS-600, Labsphere Inc.) and associated with the PD current that was simultaneously measured with another SMU (SMU-6485, Keithley). For the luminance scan where only the PD current was measured, a linear relationship between luminance and PD current is assumed. The sampling interval was 1 s.

Lifetime: The LEC's lifetime was measured with an OLED-lifetime measurement setup developed by Novaled GmbH. This setup consists of separate measurement slots on a rack. One measurement slot consists of a sample holder with electrical contacts as well as an optical detector unit with RGB photodiodes. All four pixels can be controlled separately, and voltages up to 12 V can be applied. The measurement parameters were controlled by a computer program, which also processes and stores the acquired data. The time interval in which measurement data is taken is set to 10 minutes. The luminance calibration of the detector unit was done with a red and green OLED by measuring their luminance at three different current densities with a luminance meter (LS-110, Konica Minolta) and measuring the photocurrent of the photodiodes at these respective current densities. A calibration function $L(I_{ph})$ was set up for the red and green photodiodes. Thus, for the yellow-emitting LECs, a luminance value was obtained from both the red and the green calibration. An average value was then calculated from these.

The EL spectra were measured with a calibrated spectrometer (CDS-600, Labsphere Inc.) in a calibrated integrating sphere (LMS-100, Labsphere Inc.)

*Lab G*

Ink fabrication

After dissolving $KCF_3SO_3$ in cyclohexanone for 24 h and cooling down to room temperature, the solution was filtered using a 0.45 µm PTFE filter (VWR No. 76479-008), we did not use a 0.1 µm filter.

Device fabrication

The LECs were fabricated in an argon-filled glovebox ([O2]<7 ppm, [H2O]<2 ppm) as prescribed. The AM ink was at room temperature when spin-coated. We used ITO/glass substrates (from Geomatec) with d(ITO) = 140 nm, ≈ 11 Ω/sq, T@550 nm ≈89%. We fabricated devices with an active area of 3.1 mm$^2$. We fabricated two batches of 4 devices, and we measured two cells on each device, the first cell immediately after fabrication, the second after 24 h.

Device characterization

The devices were electrically driven and measured by a computer-controlled source-measure unit (Keithley 2400), with the voltage compliance set to 21 V. Non-encapsulated devices were placed in an airtight holder, and the luminance was measured under argon atmosphere outside the glove box at room temperature using a Konica Minolta luminance meter with a close-up lens 110 at a detection angle of 0°. We used two measurement setups in parallel: setup 1, luminance meter LS110, measuring distance 205 mm, measuring diameter 0.5 mm; setup 2, LS100, measuring distance 205 mm, measuring diameter 1.5 mm. Samples were not exposed to air before or during the measurement. Sampling rates were 0.6 s$^{-1}$ for setup 1, and 0.3 s$^{-1}$ for setup 2.

The layer thickness was measured with a Veeco Dektak 6M Stylus profilometer.

EL spectra using an integrating sphere were measured on an Ocean Optics QEPro. We used a fresh cell and stopped the J-V measurement after 1 min, 1 h and 20 h for measuring the EL.

*Lab H*

Ink fabrication

The AM ink was prepared in a $N_2$-filled glovebox ([O2], [H2O] < 3 ppm) according to the attached recipe document. TMPE-OH and $KCF_3SO_3$ were dried on a hotplate in the glovebox due to the lack of a vacuum oven.

Device fabrication

The LECs were fabricated in an $N_2$-filled glovebox ([O2] < 0.1 ppm, [H2O] < 1 ppm, T ≈ 27 – 33 °C), connected to the glovebox line (measurement box and evaporator) through the inter-glovebox transfer chambers. Material specifics: Indium tin oxide (ITO)-coated (thickness = 145 nm, Rs = 20 $\Omega\ sq^{-1}$) glass substrates (substrate area = 20 x 15 $mm^2$, thickness = 1.1 mm, Ossila Ltd., UK). Eight LEC emissive pixels were fabricated on each ITO-coated glass substrate, with the emissive pixel area being 2 x 2 $mm^2$.

Device characterization

The AM thickness ($d_{AM}$) was measured with a color 3D laser microscope (VK9710K, Keyence, USA) by scratching the AM layer in the sample middle and measuring it at six different positions. The voltage-luminance transients were recorded on non-encapsulated devices inside a $N_2$-filled measurement glovebox ([O2] ≈ 1 – 3 ppm, [H2O] ≈ 2.5 – 3.5 ppm, T ≈ 26 - 30 °C). The devices that were electrically driven and the chemicals used did not get in contact with air at any time.

The LECs were electrically driven and measured by a computer-controlled source-measure unit (B2909A, Keysight, USA), with the voltage compliance set to 21 V. The luminance was measured with the calibrated photodiode (PD) (818-UV, Newport) connected to an optical power meter (1936-R, Newport). The photodiode was cross-calibrated with a spectroradiometer (CS-2000, Konica Minolta). The sampling interval during the measurement was ca. 1 s.

The EL spectra were measured with a spectroradiometer (CS-2000, Konica Minolta) positioned about 30 cm away from the LEC through the glovebox window (measurement angle of 0.1°), whereby the impact of the window was corrected.

*Lab I*

Ink fabrication

The $KCF_3SO_3$ solution was filtered by using a PTFE filter (0.22 µm pore size) after dissolving the salt in cyclohexanone for 24 h.

Device fabrication

Indium tin oxide (ITO)-coated (thickness = 150 nm, Rs = 15 Ω $sq^{-1}$) glass substrates (substrate area = 30 x 30 $mm^2$, thickness = 0.7 mm, Naranjo Substrates). Different pixel configurations were used to measure the electroluminescence (EL) and voltage-luminance transients, sixteen and four emissive pixels were fabricated on each ITO-coated glass substrate, with the emissive area being 0.0825 $cm^2$ and 0.06 $cm^2$, respectively.

Device characterization

The AM thickness ($d_{AM}$) was measured with a stylus profilometer (Alpha-Step® D-500.) by scratching the AM layer in the sample top, middle, and bottom and measuring it at 5 different positions. The voltage-luminance transients were recorded on non-encapsulated devices inside a $N_2$-filled glovebox ($[O_2] < 1$ ppm, $[H_2O] < 1$ ppm, T ≈ 26 °C). The LECs were electrically driven, and voltage and luminance versus time were monitored by using a True Color Sensor MAZeT (MTCSiCT sensor) with a Botest OLT OLED Lifetime-Test system with the compliance set to 20 V. The distance between the sample and photodetector is 7 mm. Sampling interval during the whole measurement is 1s. EL spectra were measured with an Avantes AvaSpec-2048L spectrometer with a fiber optic (FC-UVIR600-2, core size 600 µm, SMA terminations) positioned about 4 mm away from the LEC. Note: Devices were not exposed to air before nor during the characterization.

**Surface temperature of hot plate**

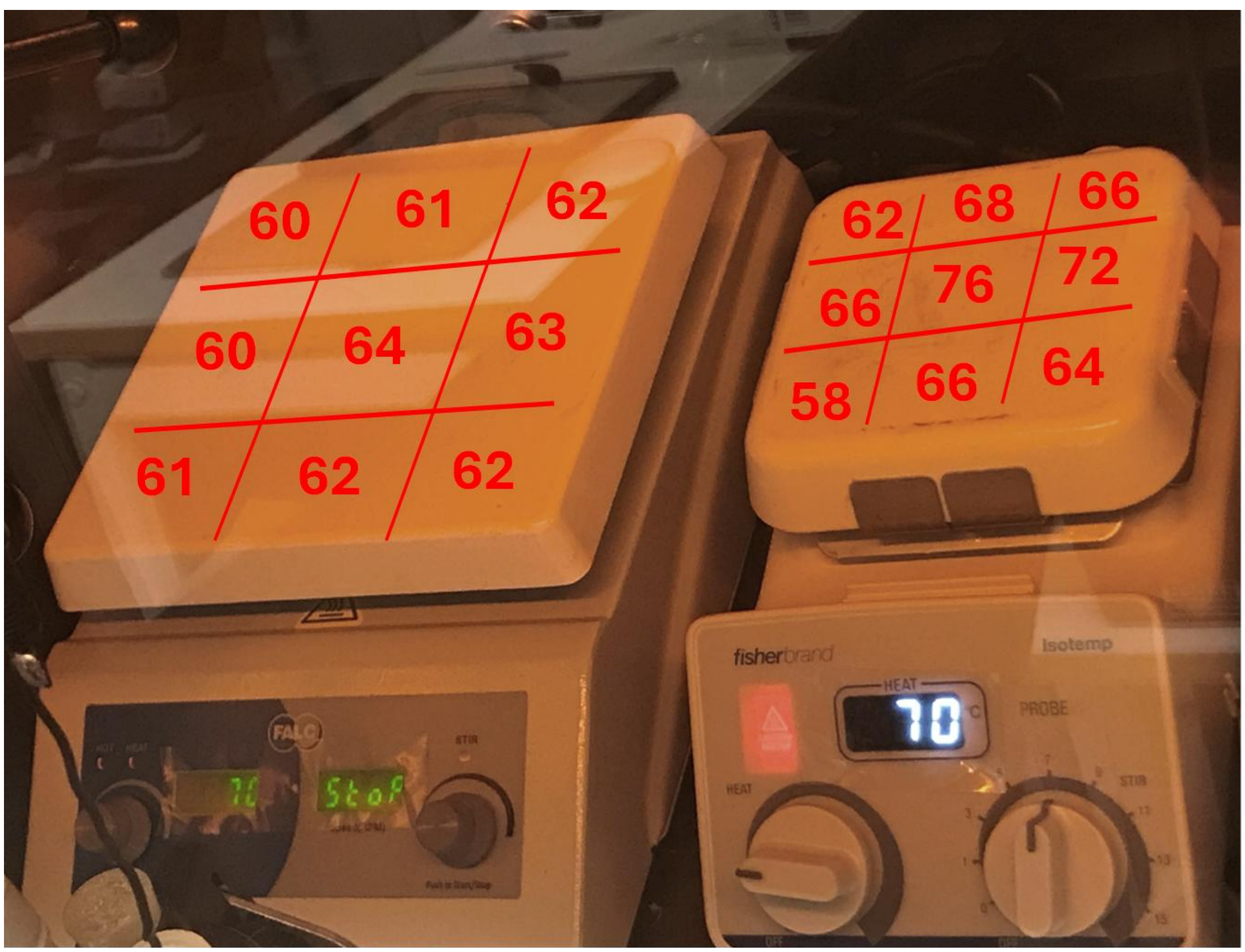

***Figure S1****. Surface temperature variation of two hotplates set to 70 °C. The temperature was measured by first attaching black tape onto the hotplate surface (to avoid light reflection) and then measuring the temperature with the aid of an infrared thermometer (Traceable, Fisher Scientific).*

For the annealing of the spin-coated AM films, the true surface temperature of the hot plate is decisive, as it influences the properties of the AM film and thereby the LEC performance. Figure S1 illustrates how the measured (true) surface temperature of a hot plate may differ from the set temperature. For consistent LEC fabrication, it is thus crucial to set the hot-plate temperature to a value that generates the required surface temperature (here, 70 °C) at the spot where the AM film is dried.

**LEC electroluminescence spectra**

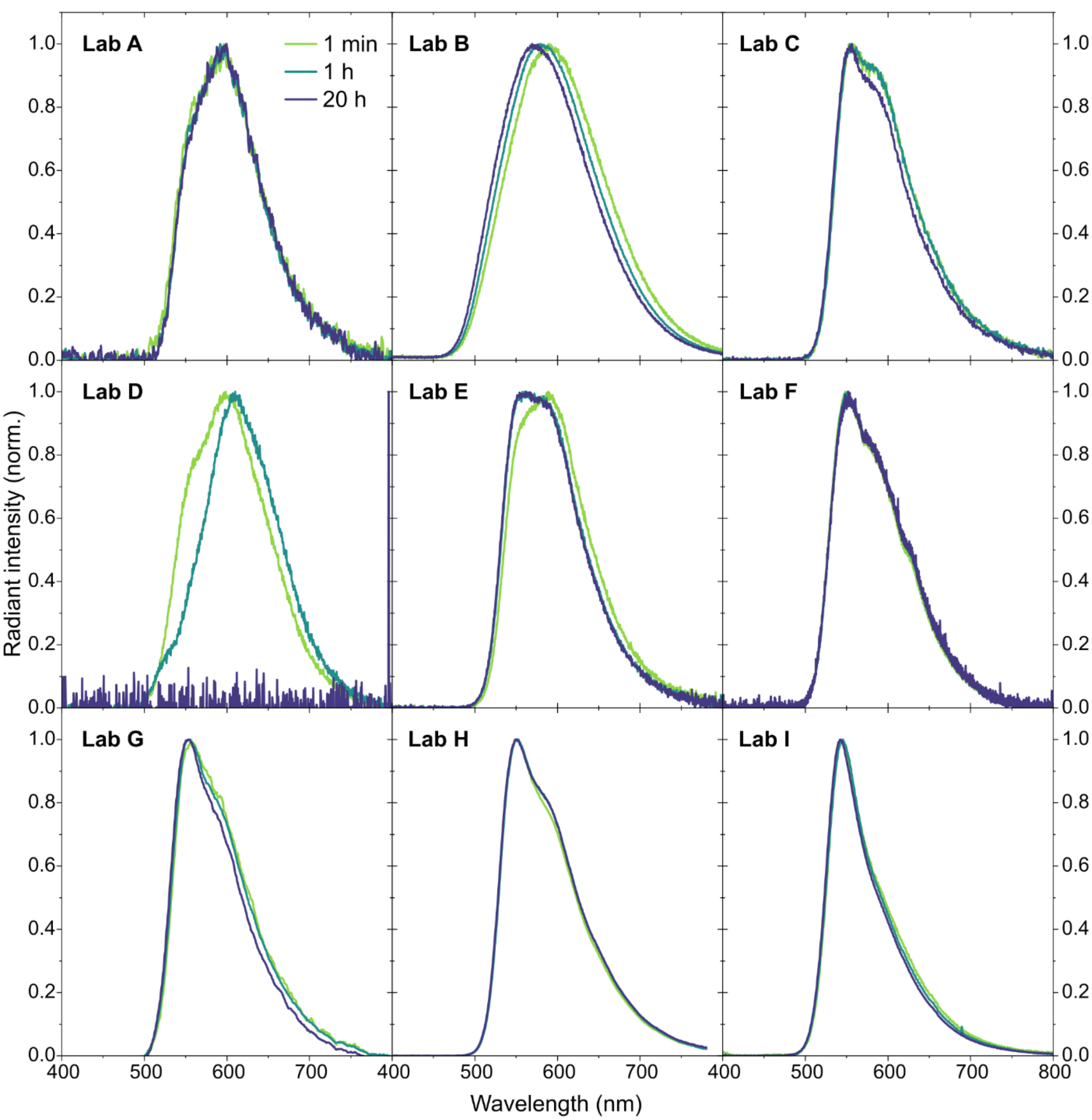

***Figure S2.*** *The electroluminescence spectra of LEC devices from each research group ("Lab"), as recorded after 1 min, 1 h, and 20 h of driving at a constant current density of 10 mA cm*$^{-2}$.

**Angle-dependent emission intensity of the reference LEC**

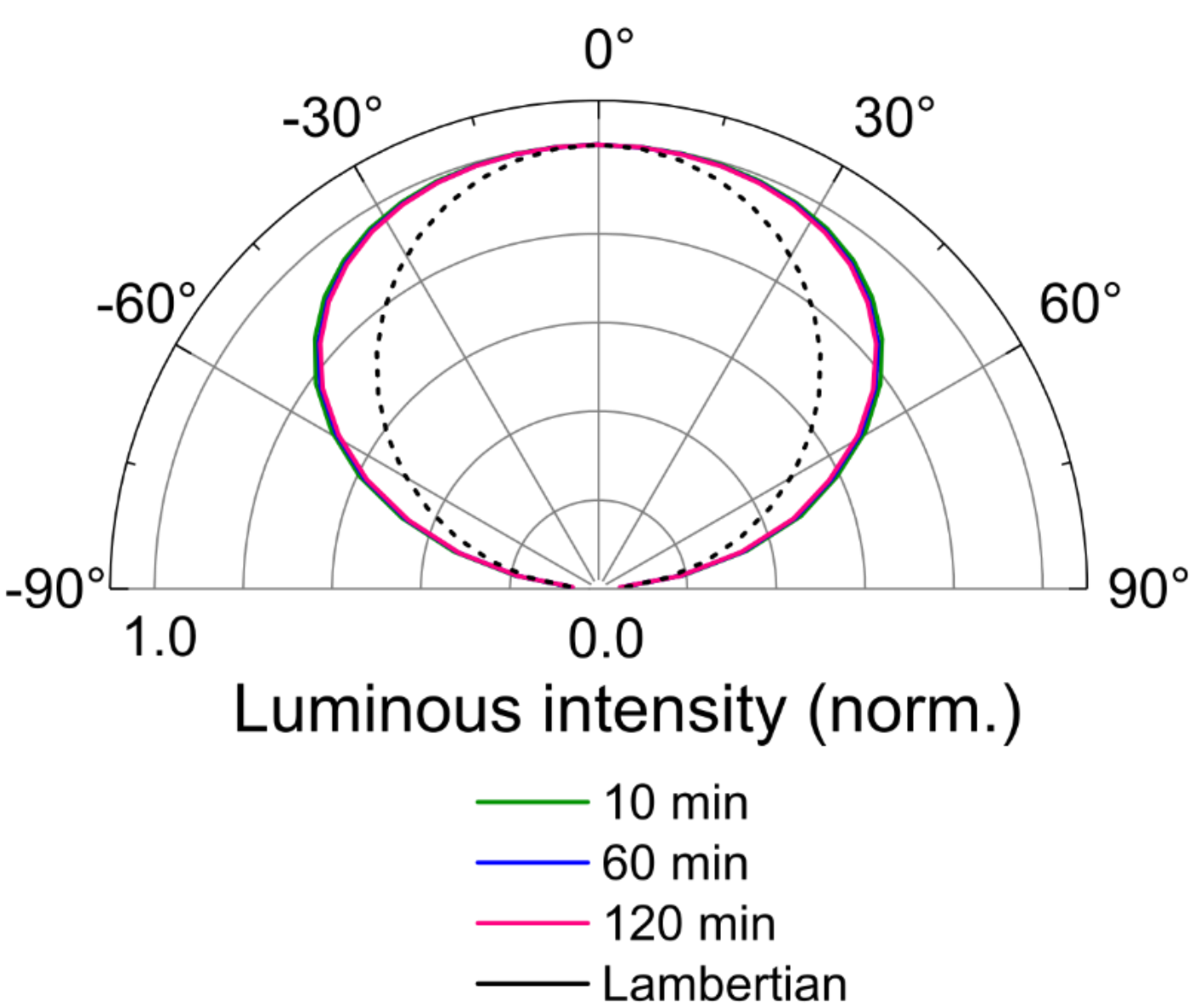

***Figure S3.*** *The normalized measured luminous intensity as a function of the viewing angle of a pristine LEC, following electric driving for the time indicated in the lower inset. The dashed black line represents the behavior of a Lambertian emitter.*

The angle-resolved EL intensity of the reference LEC was measured with a custom-built, automated spectrogoniometer setup. The device under study was placed in a connection jig, which aligned the LEC emission area with the rotation axis of a stepper motor. This rotation defined the viewing angle of the device, with 0° corresponding to the forward emission. The viewing angle was varied between −86° and 86° in steps of 5°, and this rotation was controlled by a Python-based virtual instrument. A fraction of the device emission was collected by a collimating lens (Ø = 7.2 mm, F230 SMA-A, Thorlabs, Germany) positioned 75 mm away from the device, resulting in a small and constant solid collection angle (Ω) of 0.007 sr. An optical fiber delivered the collected light to a CCD-array spectrometer (Flame-S, OceanOptics). More details on the measurement setup and procedure can be found in [E. M. Lindh, P. Lundberg, T. Lanz, J. Mindemark, L. Edman, *Sci. Rep.* 2018, 8, 6970].

**Forward luminance as a function of AM thickness**

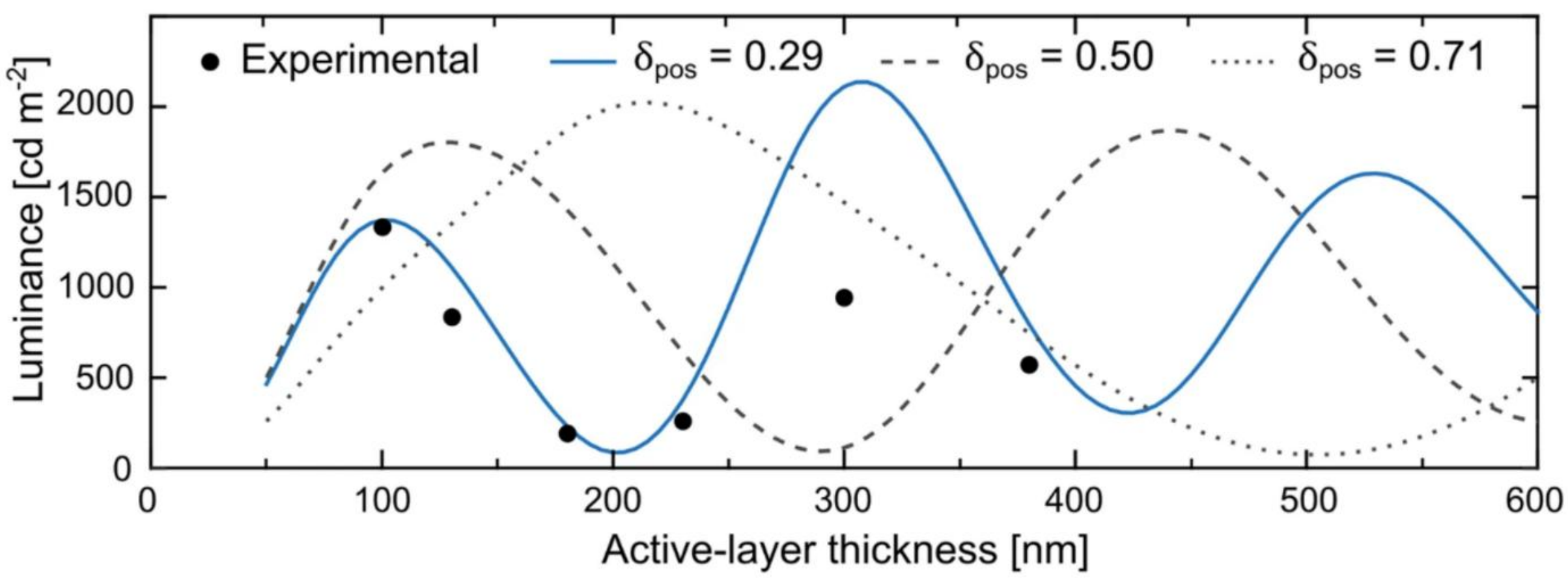

***Figure S4.*** *The measured (circles) and the simulated (lines) LEC forward luminance as a function of active-layer (or AM) thickness during driving by a constant current density of 25 mA cm*$^{-2}$*.* $\delta_{pos}$ *indicates the relative emission zone position in the AM, with 0 being the ITO (anode) interface and 1 the Al (cathode) interface. [Reprinted from Lindh et al., Sci Rep 9, 10433 (2019) as permitted by CCBY 4.0].*